\documentclass[aps,prb,twocolumn,eqsecnum,floatfix,showpacs,showkeys]{revtex4}
\usepackage{amsmath,dcolumn}
\usepackage{mathrsfs}
\usepackage{bbold}
\makeatletter
\def\bra#1{\mathinner{\langle{#1}|}}
\def\ket#1{\mathinner{|{#1}\rangle}}
\def\braket#1{\mathinner{\langle{#1}\rangle}}

{\catcode`\|=\active 
  \gdef\Braket#1{\left<\mathcode`\|"8000\let|\BraVert {#1}\right>}}
\def\BraVert{\egroup\,\mid@vertical\,\bgroup}
{\catcode`\|=\active
  \gdef\set#1{\mathinner{\lbrace\,{\mathcode`\|"8000\let|\midvert #1}\,\rbrace}}
  \gdef\Set#1{\left\{\:{\mathcode`\|"8000\let|\SetVert #1}\:\right\}}}
\def\midvert{\egroup\mid\bgroup}
\def\SetVert{\egroup\;\mid@vertical\;\bgroup}
\begingroup
 \edef\@tempa{\meaning\middle}
 \edef\@tempb{\string\middle}
\expandafter \endgroup \ifx\@tempa\@tempb
 \def\mid@vertical{\middle|}
\else
 \let\mid@vertical\vrule
\fi
\makeatother

\newcommand{\cx}[1]{\smash{c_{#1}^{\phantom{\dagger}}}}
\newcommand{\cxd}[1]{\smash{c_{#1}^{\dagger}}}
\newcommand{\ck}[1]{{\smash{\tilde{c}_{#1}^{\phantom{\dagger}}}}}
\newcommand{\ckd}[1]{\smash{\tilde{c}_{#1}^{\dagger}}}
\newcommand{\expn}[3]{\ensuremath{\braket{#1|#2|#3}}}
\newcommand{\inprod}[2]{\ensuremath{\braket{#1|#2}}}
\newcommand{\mean}[1]{\ensuremath{\braket{#1}}}
\newcommand{\nbar}{\ensuremath{\bar{n}}}
\DeclareMathOperator{\sgn}{sgn}
\DeclareMathOperator{\rank}{rank}

\DeclareMathOperator{\diag}{diag}
\DeclareMathOperator{\Tr}{Tr}
\newcommand{\one}{\ensuremath{\mathbb{1}}}
\newcommand{\zero}{\ensuremath{\mathbb{0}}}
\newcommand{\smallone}{\one}

\begin{document}
\title{Many-Body Density Matrices for Free Fermions}
\author{Siew-Ann \surname{Cheong}}
\author{Christopher L. \surname{Henley}}
\affiliation{Laboratory of Atomic and Solid State Physics, Cornell University}
\date{\today}

\begin{abstract}
\bigskip
Building upon an analytical technique introduced by Chung and Peschel,
Phys. Rev. B \textbf{64}, art. 064412 (2001), we calculated the many-body
density matrix $\rho_B$ of a finite block of $B$ sites within an infinite 
system of free spinless fermions in arbitrary dimensions.  In terms of the 
block Green function matrix $G$ (whose elements are $G_{\bar ij} = 
\braket{\cxd{i}\cx{j}}$, where $\cxd{i}$ and $\cx{j}$ are fermion creation 
and annihilation operators acting on sites $i$ and $j$ within the block 
respectively), the density matrix can be written as $\rho_B = \det(\one - G)
\exp[\sum_{ij}(\log G(\one - G)^{-1})_{ij}\cxd{i}\cx{j}]$.  Our results
suggests that Hilbert space truncation schemes should retain the states 
created by a subset of the $\cxd{i}$'s (in any combination), rather than
selecting eigenvectors of $\rho_B$ independently based on the eigenvalue.
\end{abstract}



\maketitle

\section{Introduction}

Exact solutions are hard to come by in many-body problems, and every so often
we have to resort to numerical solutions.  The traditional approaches, applied
to finite systems, are exact diagonalization and quantum Monte Carlo.  For 
quantum lattice models of fermions, the former is constrained by the size of
the Hilbert space, which grows exponentially with the number of sites, while 
the latter is plagued by the `minus-sign problem'.  For quantum lattice models 
of bosons, the Hilbert space is infinite-dimensional even for finite systems.
In either case, because of the enormous computational complexity involved, 
there is no hope of getting to the thermodynamic limit of 
infinite system size.  In view of these difficulties, one then hopes for the 
next best thing: approximate solutions that captures the essence of the physics.

This is where renormalization group (RG) approaches comes in.  In 
such approaches\cite{burkhardt82,zivkovic90,cheranovski94,malrieu01,%
lepetit93,monthoux96,monthoux97,degenhard00,bursill99,white92,white93}
to the approximate solution of otherwise intractable problems, 
the size of the Hilbert space is kept in check by aggressive truncation, with 
the hope that the small number of states kept will reproduce the more important 
features of the physics.  Whatever the RG scheme, ultimately its success will 
lie in how the truncation is done.  Since the quantum-mechanical state of a
block embedded in a larger system must in general be described by a density 
matrix, it is therefore natural to use it to guide the truncations.

With the goal of understanding the structure and spectrum of the density 
matrix, and their implications on RG truncation schemes in mind, Peschel 
\emph{et al} calculated exactly the half-chain density matrix for several
models.\cite{peschel99a,peschel99b,chung01} For a chain of coupled harmonic 
oscillators and spinless Bogoliubov fermions, they found that the half-chain 
density matrices can be expressed exactly as the exponential of a 
pseudo-Hamiltonian, whose spectrum is generated by a set of independent 
bosonic and fermionic operators respectively.  In this paper, we pursue their 
analysis further for a system of free spinless fermions to obtain a 
closed-form formula relating the density matrix $\rho_B$ of a subsystem and
the subsystem Green function matrix $G$ (to be defined in 
Section~\ref{sect:index}).

The organization of the paper will be as follows: we will start in 
Section~\ref{sect:dmblock} by reviewing the density matrix formulation of 
quantum mechanics, and how the density matrix $\rho_B$ of a subsystem can be 
obtained from the density matrix $\rho_0$ of the overall system.  Following 
this, we will describe an alternative approach to calculating the density 
matrix elements as expectations of referencing operators.  We shall show that 
the real-space structure, and the strong signs that point to a closed-form 
expression for $\rho_B$,  is most readily discerned within this alternative 
formulation.  Then, in Section~\ref{sect:exact}, we derive this closed-form 
expression for $\rho_B$ in terms of the subsystem Green function matrix $G$ 
by adapting the technique put forward by Chung and Peschel.\cite{chung01}  
The existence of such a relation between $\rho_B$ and $G$ tells us that 
$\rho_B$ is completely determined by its 0- and 1-particle
sectors.  We discuss the implications of this in Section~\ref{sect:checks}, 
where we illustrate how the eigenvectors and eigenvalues of the 
$(F > 1)$-sectors of $\rho_B$ can be constructed out of the eigenvectors and 
eigenvalues of the 1-particle sector.  We also show how symmetries of the 
Hamiltonian that are realized in $\rho_B$ affect the pattern of degeneracies 
in the eigenvalues of these sectors, an understanding of which is important 
in formulating a consistent truncation scheme.

\section{Density Matrix of a Finite Block}
\label{sect:dmblock}

In this section, we first review the density matrix notions that
will be used throughout this paper.  Following this we develop the first
of our two paths to calculate the density matrix $\rho_B$ for a particular 
block within a large system of non-interacting fermions.  By analyzing the
index structure of the matrices involved, we arrived at a conjecture for a
closed-form expression for the 1-particle sector of the block density matrix
$\rho_B$ in terms of the block Green function matrix $G$. 

\subsection{Density Matrix Formulation of Quantum Mechanics}
\label{sect:revdenmat}

In quantum mechanics one distinguishes between \emph{pure states}, which 
occur, for example, at $T = 0$ when the system is totally decoupled from the 
rest of the universe, and \emph{mixed states}, which occur, for example, at 
$T > 0$ when the system is in thermodynamic equilibrium with the rest of the 
universe.  A pure state can be described by a wavefunction $\ket{\Psi}$ in the 
usual formulation of quantum mechanics, whereas a mixed state cannot.  Both 
type of states are treated on equal footing in the density matrix formulation 
of quantum mechanics, in which the state of a system is described by a density 
matrix $\rho_0$ (see, for example, Ref.~\onlinecite{ballentine98}).  In this
formulation of quantum mechanics, the expectation of an operator $A$ in a state
described by $\rho_0$ is given by
\begin{equation}\label{eqn:rhonought}
\mean{A} = \Tr (\rho_0 A).
\end{equation}
If the state so described is pure, i.e.~given in the usual formulation by
the wavefunction $\ket{\Psi}$, so that $\mean{A} = \braket{\Psi | A |\Psi}$,
then it is clear that $\rho_0 = \ket{\Psi}\bra{\Psi}$.

In this paper, we shall be mainly interested in a finite subsystem of $B$ 
sites, which we call the \emph{block}, embedded within a larger system of $N$
sites, with periodic boundary conditions in $d$ dimensions.  The latter can 
then be taken to the thermodynamic 
limit of infinite number of sites, i.e.~$N \to \infty$.  The system minus the 
block is called the \emph{environment} of the
block.  If the overall system is known to be in a pure state $\ket{\Psi}$,
then in general the quantum-mechanical state of the block cannot be described 
by a pure state wavefunction.  Instead, the mixed state of the block must be 
described by a block density matrix $\rho_B$ (see arguments in 
Ref.~\onlinecite{feynman72}), so defined that
\begin{equation}\label{eqn:rhodefine}
\mean{A} = \Tr(\rho_B A), 
\end{equation}
if the operator $A$ acts entirely within the block.  

There are two useful formulas to relate the block density matrix $\rho_B$ to
the density matrix $\rho_0$ of the entire system.  The first formula,
which we will used in Section~\ref{sect:exact}, follows from 
\eqref{eqn:rhonought} and \eqref{eqn:rhodefine}.  Using the subscripts $B$ and 
$E$ respectively to make the trace over the degrees of freedom associated with 
the block and its environment more explicit, we can rewrite 
\eqref{eqn:rhonought} as
\begin{equation}
\mean{A} = \Tr(\rho_0 A) = \Tr_{B,E}(\rho_0 A).
\end{equation}
Since $A$ does not act on the environment, we can trivially trace over 
environmental degrees of freedom to get
\begin{equation}
\mean{A} = \Tr_B\left\{\left[\Tr_E(\rho_0)\right] A\right\}.
\end{equation}
Comparing this with \eqref{eqn:rhodefine}, we find a consistent expectation 
for $A$ whether it is taken over the entire system or just over the block,
if the block density matrix is defined as
\begin{equation}\label{eqn:rhoBrhonought}
\rho_B = \Tr_E(\rho_0).
\end{equation}

The second formula for $\rho_B$ allows us to write down its matrix elements 
explicitly when the overall system is in a pure state.  To arrive at this 
formula, let us first note that any pure state of the overall system can be 
written as $\ket{\Psi} = \sum_b \ket{b}\ket{e_b}$, 
where $\ket{b}$ is a complete orthonormal (many-body) basis for the block, 
and $\ket{e_b}$ is the (unnormalized) state of the environment associated with 
the state $\ket{b}$ on the block.  Using this form for $\ket{\Psi}$ in
\eqref{eqn:rhonought}, we find that
\begin{equation}
\mean{A} = \sum_{b,b'} \bra{e_b}\bra{b} A \ket{b'}\ket{e_{b'}} =
\Tr_B(\rho_B A)
\end{equation}
if the block density matrix $\rho_B$ is defined such that
\begin{equation}\label{eqn:second}
(\rho_B)_{b'b} = \inprod{e_b}{e_{b'}},
\end{equation}
i.e.~the matrix element of $\rho_B$ between $\ket{b}$ and $\ket{b'}$ is
none other than the overlap between their associated environmental states 
$\ket{e_b}$ and $\ket{e_{b'}}$.

\subsection{Free Spinless Fermions}
\label{sect:freespinlessfermions}

Let us now apply \eqref{eqn:second} to calculate the block density matrix
from the ground state of a ring of $N \to \infty$ free spinless fermions, 
the simplest realization of which is described by a translationally-invariant 
Hamiltonian with nearest-neighbor hopping
\begin{equation}\label{eqn:hamiltonian}
H = -t\sum_{\langle ij\rangle}
\left(\cxd{i}\cx{j} + \cxd{j}\cx{i}\right),
\end{equation}
where $\cx{i}$ and $\cxd{i}$ are the fermion annihilation and creation 
operators acting on site $i$, and $\langle ij\rangle$ runs once
over each pair of neighbor sites.  

The Hamiltonian given in \eqref{eqn:hamiltonian} is diagonal in momentum
space, and can be written as
\begin{equation}\label{eqn:hamfourier}
H = \sum_k \epsilon_k \ckd{k} \ck{k}. 
\end{equation}
Here
\begin{equation}
\begin{aligned}
\ck{k} &\equiv N^{-1/2}\sum_j\cx{j}\,e^{ik\cdot r_j}, \\
\ckd{k} &\equiv N^{-1/2}\sum_j\cxd{j}\,e^{-ik\cdot r_j}
\end{aligned}
\end{equation}
are the momentum space annihilation and creation operators, $r_i$ is the 
position of site $i$, and $\epsilon_k$ the single-particle energy associated 
with wavevector $k$.  The ground state of such a system is just a Fermi sea 
\begin{equation}\label{eqn:fermisea}
\smash{\ket{\Psi_F} = \prod_{\text{$k$ filled}} \ckd{k}\ket{0}}, 
\end{equation}
where $\ket{0}$ is the vacuum, and the product is over the wavevectors
inside the Fermi surface.

As noted in \eqref{eqn:second}, when the ground state wavefunction is written
as $\ket{\Psi_F} = \sum_b \ket{b}\ket{e_b}$, the block density matrix elements 
are $(\rho_B)_{b'b} = \inprod{e_b}{e_{b'}}$.  When dealing with a finite block
and an infinite environment, it makes no sense to evaluate these environmental
overlaps by first calculating $\ket{e_b}$ and $\ket{e_{b'}}$.  Instead,
we find that it possible to evaluate such environmental overlaps with the
help of operator products that are defined entirely within the block.  To do
so, let us first write the many-body states $\ket{b}$ on the block in the 
occupation number representation as $\ket{b} = \ket{n_1^b n_2^b \cdots n_B^b}$,
where $n_j^b = 0$ or 1 depending on whether the site $j$ on the block
is empty or occupied in the state $\ket{b}$.  We then define the
\emph{referencing operators}
\begin{equation}\label{eqn:Koperators}
K_b = \prod_{j = 1}^B\left[n_j^b \cx{j} + (1 - n_j^b)\cx{j}\cxd{j}\right],
\end{equation}
such that the effect of $K_b$ acting on a state $\ket{b'}$ is 
$K_b\ket{b'} = \delta_{bb'}\ket{0}_B$, where $\ket{0}_B$ is the 
\emph{reference state} for which all sites on the block are empty.
Letting $K_b$ act on $\ket{\Psi_F}$ gives
$K_b\ket{\Psi_F} = \sum_{b''} K_b\ket{b''}\ket{e_{b''}} = \ket{0}_B\ket{e_b}$.
Hence, in terms of the operators $K_b$ and their conjugates $K_b^{\dagger}$,
the density matrix elements are found to be
\begin{equation}\label{eqn:rhowork}
(\rho_B)_{b'b} = \inprod{e_b}{e_{b'}} = 
\expn{\Psi_F}{K_b^{\dagger} K_{b'}}{\Psi_F} = \mean{K_b^{\dagger}K_{b'}}.
\end{equation} 

From the way the operators $K_b$ are defined, we know that $\rho_B$ is real and 
symmetric.  Furthermore, $(\rho_B)_{b'b}$ vanishes if the states $\ket{b}$ and 
$\ket{b'}$ do not contain the same number of fermions $F$.  Consequently, the 
non-zero matrix elements of $\rho_B$ are found in a total of $(B + 1)$ 
submatrices along the diagonal, corresponding to the various $F$-particle 
sectors, for $F = 0$, 1, \dots, $B$.  We shall call such submatrices 
$\rho_{B,F}$, and their eigenvalues the \emph{density matrix weights} 
$w_{B,F,l}$, where $l = 1, \dots, \rank(\rho_{B,F})$.

\subsection{Conjecture Based on Index Structure}
\label{sect:index}

In general, for a block of $B$ sites, there are a total of $2^{B}$ $K_b$
operators we need to write down explicitly to calculate the ${\sim}2^{B}$ 
density matrix elements.  For large blocks, this is extremely tedious and 
has to be automated (see Appendix \ref{app:automatic}), but for small blocks, 
it is not difficult to work out exact expressions for $(\rho_B)_{b'b}$ in 
terms of the $2n$-point functions
\begin{subequations}\label{eqn:twonptfunctions}
\begin{align}
G_{\bar ij} &\equiv \mean{\cxd{i}\cx{j}}, \\
G_{\bar i\bar j kl} &= \mean{\cxd{i}\cxd{j}\cx{k}\cx{l}} =
(-1)^{\frac{2(2-1)}{2}}\begin{vmatrix}
G_{\bar ik} & G_{\bar il} \\
G_{\bar jk} & G_{\bar jl} \end{vmatrix}, \\
G_{\bar i\bar j\bar k lmn} &\equiv 
\mean{\cxd{i}\cxd{j}\cxd{k}\cx{l}\cx{m}\cx{n}} \notag \\
&= (-1)^{\frac{3(3-1)}{2}}\begin{vmatrix}
G_{\bar il} & G_{\bar im} & G_{\bar in} \\
G_{\bar jl} & G_{\bar jm} & G_{\bar jn} \\
G_{\bar kl} & G_{\bar km} & G_{\bar kn} \end{vmatrix},\ 
\end{align}
\end{subequations}
and so forth, where $i,j, k, l, \dots = 1, \dots, B$ are sites on the block.
As shown explicitly above, the $2n$-point functions $G_{\bar i_1\cdots\bar i_n 
j_1\cdots j_n}$ Wick factorizes into sums of products of 2-point functions 
$G_{\bar i j}$ for our non-interacting system, with an overall fermion
factor of $(-1)^{n(n-1)/2}$.  

At this point let us note that since the 2-point functions $G_{\bar ij}$ are 
labelled by two indices, it is convenient to organize them into a \emph{system 
Green function matrix} $\mathscr{G}$ given by
\begin{equation}\label{eqn:systemgreenfunctionmatrix}
\mathscr{G} = \begin{bmatrix}
G_{\bar 11} & \cdots & G_{\bar 1 B} & G_{\bar 1 B+1} & \cdots & G_{\bar 1 N} \\
G_{\bar 21} & \cdots & G_{\bar 2 B} & G_{\bar 2 B+1} & \cdots & G_{\bar 2 N} \\
\vdots & \ddots & \vdots & \vdots & \ddots & \vdots \\
G_{\overline{B} 1} & \cdots & G_{\overline{B} B} &
G_{\overline{B} B+1} & \cdots & G_{\overline{B} N} \\
G_{\overline{B+1} 1} & \cdots & G_{\overline{B+1} B} &
G_{\overline{B+1} B+1} & \cdots & G_{\overline{B+1} N} \\
\vdots & \ddots & \vdots & \vdots & \ddots & \vdots \\
G_{\overline{N} 1} & \cdots & G_{\overline{N} B} &
G_{\overline{N} B+1} & \cdots & G_{\overline{N} N} \end{bmatrix},
\end{equation}
of which
\begin{equation}\label{eqn:blockgreenfunctionmatrix}
G = \begin{bmatrix}
G_{\bar 11} & G_{\bar 12} & \cdots & G_{\bar 1 B} \\
G_{\bar 21} & G_{\bar 22} & \cdots & G_{\bar 2 B} \\
\vdots & \vdots & \ddots & \vdots \\
G_{\overline{B} 1} & G_{\overline{B} 2} & \cdots & G_{\overline{B} B} 
\end{bmatrix}
\end{equation}
is its \emph{restriction to the block}.  We call $G$ the \emph{block Green 
function matrix}.  As a result of the translational invariance of $H$, 
$\mathscr{G}$ is also translationally invariant.  In real space, this 
means that its matrix elements $\mathscr{G}_{ij} = G_{\bar ij} = 
\braket{\cxd{i}\cx{j}}$ are functions only of $r_i - r_j$.  When $\mathscr{G}$ 
is restricted to the block to give $G$, however, this translational invariance 
is lost due to the fact that the presence of a block in the system allows an 
unambiguous definition of the origin.

Anyway, from \eqref{eqn:Koperators} and \eqref{eqn:rhowork}, we see on the
one hand that $(\rho_B)_{bb'}$ can be written as sums of $2n$-point functions
--- which themselves factor into sums of products of 2-point
functions --- and so we find that $(\rho_B)_{bb'}$ are all functions of 
$G_{\bar k l}$.  On the other hand, the 1-particle sector of $\rho_B$ contains 
matrix elements $(\rho_B)_{bb'}$ connecting the states $\ket{b}$ and 
$\ket{b'}$, which contain one particle each at sites, say, $i$ and 
$j$ respectively.  Therefore, the matrix elements within $\rho_{B,1}$ may be 
indexed using $i$ and $j$ instead of $b$ and $b'$.  Diligently writing down 
the polynomial expressions
\begin{equation}
\begin{split}
(\rho_{B,1})_{ij} &= 
\sum_{k_1,l_1}^B \alpha_{ij; k_1l_1}^{(1)} G_{\bar{k}_1 l_1} + {} \\
&\quad \sum_{\substack{k_1,k_2,\\ l_1,l_2}}^B \alpha_{ij; k_1 k_2 l_1 l_2}^{(2)}
G_{\bar{k}_1 l_1} G_{\bar{k}_2 l_2} + \cdots + \\
&\quad \sum_{\substack{k_1,\dots,k_B,\\ l_1,\dots,l_B}}^B 
\alpha_{ij;k_1 \cdots k_B l_1 \cdots l_B}^{(B)}
G_{\bar{k}_1 l_1} \times \cdots \times G_{\bar{k}_B l_B},
\end{split}
\end{equation}
we find that: (a) the coefficients $\smash{\alpha_{ij ; k_1 \cdots k_n 
l_1 \cdots l_n}^{(n)}}$ are independent of $i$ and $j$; and (b) indices other 
than $i$ and $j$ always appear in pairs, as if they are summed over.  

Exhaustively comparing the matrix elements of $\rho_{B,1}$ and 
powers of $G$ for $2 \leq B \leq 5$, we find that
\begin{equation}\label{eqn:Gseries}
\begin{split}
\rho_{B,1} &= G + G^2 - G\Tr(G) + {} \\
&\quad\ G^3 - G^2\Tr(G) - \tfrac{1}{2}\left\{
\Tr(G^2) - [\Tr(G)]^2\right\} G + {} \\
&\quad\ G^4 - G^3\Tr(G) - \tfrac{1}{2}\left\{
\Tr(G^2) - [\Tr(G)]^2\right\} G^2 - {} \\
&\quad\ \left\{\tfrac{1}{3}\Tr(G^3) - \tfrac{1}{2}
\Tr(G)\Tr(G^2) + \tfrac{1}{6}[\Tr(G)^3]\right\} G + \cdots.
\end{split}
\end{equation}
What is most fascinating about this series is that for $B = 2$, 
\eqref{eqn:Koperators} and \eqref{eqn:rhowork} tell us that $\rho_{B,1}$ can
be at most $O(G^2)$, since its matrix elements never contain terms with
more than two creation and annihilation operators each.  Yet 
\eqref{eqn:Gseries} is perfectly valid for $B = 2$, because terms higher
order in $G$ vanish.  For $B = 3$ and $B = 4$, we find similarly that 
terms higher order than $O(G^3)$ and $O(G^4)$ vanishes, respectively.
If we conjecture that \eqref{eqn:Gseries} gives the leading terms to an 
infinite series that holds true for all $B > 5$, then we can factorize it into
\begin{multline}\label{eqn:Gfactorized}
\rho_{B,1} = (G + G^2 + G^3 + \cdots) \times {} \\
\exp\left[-\Tr(G + \tfrac{1}{2}G^2 + \tfrac{1}{3} G^3 + \cdots)\right].
\end{multline}
Noting that the series inside the trace is just $-\log(\one-G)$, 
\eqref{eqn:Gfactorized} can be compactly written as
\begin{equation}\label{eqn:closedform}
\rho_{B,1} = G(\one - G)^{-1}\det(\one - G).
\end{equation}

\section{Derivation and Properties of $\boldsymbol{\rho_B}$}
\label{sect:exact}

In passing from \eqref{eqn:Gseries} to \eqref{eqn:closedform}, a leap of 
faith was required, and it would appear forbiddingly difficult to actually 
prove \eqref{eqn:closedform} for arbitrary block sizes $B$, by the algebraic 
manipulations used in Section~\ref{sect:index}.  Fortunately, an alternate 
technique introduced by Chung and Peschel\cite{chung01} can be adapted and 
extended for calculating the density matrix of a finite block, although
it comes with its own set of technical difficulties.
It turns out that if the whole system were in the Fermi sea ground state, 
the derivation would require the inversion of singular matrices.  In the end,
the singularities do cancel and give a well-defined answer, but a 
regularization is needed to avoid divergences in the intermediate steps.  The 
most natural way to do so would be to generalize our problem to nonzero 
temperature, in which case the limit $T \to 0$ then provides the needed 
regularization.\footnote{Ref.~\protect\onlinecite{chung01} avoid the 
singularities by assuming a Hamiltonian with nonzero anomalous terms 
containing double creation or double annihilation operators.
Alternatively, realizing that we have definite occupation
numbers, i.e.~$\mean{\ckd{k}\tilde{c}_k^{}} = 0, 1$ at $T = 0$, 
the density matrix $\rho_0$ must be written as a product of projection 
operators, i.e.~$\rho_0 = \prod_{|k|<k_F}\ckd{k}\tilde{c}_k^{}\prod_{|k'|>k_F}
\tilde{c}_{k'}^{}\tilde{c}_{k'}^{\dagger}$.  This is possible only if 
$\tilde{\Gamma}_{kk} = +\infty$ for $|k| < k_F$ and $\tilde{\Gamma}_{kk} = 
-\infty$ for $|k| > k_F$.  For the purpose of algebraic manipulations, this 
choice of $\tilde{\Gamma}_{kk}$ must be regularized, i.e.~take 
$\tilde{\Gamma}_{kk} = \Lambda\sgn(k_F - |k|)$,
and take $\Lambda \to \infty$ at the end of the calculations.
With this choice of regularization, $e^{\Gamma}$ can then be written in
terms of the zero-temperature Green function matrix $\mathscr{G}$, whose
matrix elements in momentum space are $\mathscr{G}_{kk} = \theta(k_F - |k|)$
(where $\theta(x) = 0$ for $x < 0$ and $\theta(x) = 1$ when $x > 0$ is the
step function) as $e^{\Gamma} = e^{-\Lambda}\,\one  + (e^{\Lambda} - 
e^{-\Lambda})\,\mathscr{G}$.  It is then easy to show that $(\one + 
e^{\Gamma})^{-1} = (1 + e^{-\Lambda})^{-1}\one + [(1 + e^{\Lambda})^{-1} - 
(1 + e^{-\Lambda})^{-1}]\mathscr{G}$, which becomes $(\one - \mathscr{G})$ in 
the limit of $\Lambda \to \infty$.}

In essence, the calculations is just that of evaluating a Gaussian integral
with the usual shift in integration variables.  However, because we are dealing
with fermions, whose creation and annihilation operators anticommute rather
than commute, additional machinery is needed to accomplish the feat of
Gaussian integration.  After casting the system density matrix $\rho_0$ as
a Gaussian of the fermion operators, we introduce fermionic coherent states
with the aid of anticommuting Grassmann variables.  The matrix elements of
$\rho_0$ between such coherent states, obtained via a translation
machinery, are similarly of Gaussian form, but are now easier to handle.  A
Gaussian integration over the environmental degrees of freedom then yields
elements of the block density matrix $\rho_B$, following which reverse
translation gives $\rho_B$ proper.

\subsection{Exponential Form for \boldmath$\rho_0$}
\label{sect:expformrhonought}

To get the calculations underway, we consider the grand-canonical $T > 0$ 
density matrix $\rho_0$ of the overall system that the block is embedded in. 
As always, this is given by 
\begin{equation}\label{eqn:rho0}
\rho_0 = \mathscr{Q}^{-1} \exp [-\beta (H - \mu F)]
\end{equation}
where $\beta \equiv 1/k_BT$, $\mu$ is the chemical potential, and
$F \equiv \sum_k \ckd{k}\ck{k} = \sum_i \cxd{i}\cx{i}$ is the fermion number 
operator.  The prefactor $\mathscr{Q}^{-1}$ in \eqref{eqn:rho0} is just the 
reciprocal of the grand partition function, to ensure that $\Tr(\rho_0) = 1$. 

The notations can be made more compact if we introduce the matrices $\Gamma$
and its Fourier transform $\tilde{\Gamma}$, such that
\begin{equation}\label{eqn:rhonoughtgamma}
\rho_0 = \mathscr{Q}^{-1} \exp\left(\textstyle\sum_{i,j} 
\Gamma_{ij}\cxd{i}\cx{j}\right) = \mathscr{Q}^{-1} 
\exp\left(\textstyle\sum_k \tilde{\Gamma}_{kk}\ckd{k}\ck{k}\right),
\end{equation}
where we have made use of the fact that $H - \mu F$, and hence 
$\tilde{\Gamma}$, is diagonal in momentum space.  The matrix elements
of $\Gamma$ can be read off from \eqref{eqn:hamiltonian} as
\begin{equation}
\Gamma_{ij} = \begin{cases}
\; \beta\mu, & \text{if $i = j$;} \\
\; \beta t, & \text{if $i$ and $j$ are nearest neighbors;} \\
\; 0, & \text{otherwise,} \end{cases}
\end{equation}
while those of $\tilde{\Gamma}$ can be read off from
\eqref{eqn:hamfourier} as
\begin{equation}
\tilde{\Gamma}_{kk} = -\beta E_k,
\end{equation}
where $E_k \equiv \epsilon_k - \mu$ is the single-particle energy measured 
relative to $\mu$.

In order to prove our conjecture \eqref{eqn:closedform}, it is clear that
we need to somehow relate $\Gamma$ to $\mathscr{G}$.  To do this, let us note
that since $\mathscr{G}$ is translationally invariant, its Fourier transform
$\tilde{\mathscr{G}}$ is diagonal in momentum space, with matrix elements given
in the grand-canonical ensemble as
\begin{equation}
\tilde{\mathscr{G}}_{kk} = \mean{\ckd{k}\ck{k}} = \frac{1}{\exp\beta E_k + 1},
\end{equation}
observing which we find that
\begin{equation}
\tilde{\mathscr{G}}_{kk} = \exp(\tilde{\Gamma}_{kk})
\left[\exp(\tilde{\Gamma}_{kk}) + 1\right]^{-1}.
\end{equation}
But since both $\tilde{\mathscr{G}}$ and $\tilde{\Gamma}$ are diagonal
matrices, we have the relation
\begin{equation}\label{eqn:exptildegamma}
e^{\tilde{\Gamma}} = \tilde{\mathscr{G}}(\one - \tilde{\mathscr{G}})^{-1},
\end{equation}
where $e^{\tilde{\Gamma}}$ is the matrix exponential of $\tilde{\Gamma}$.

Of course, $\mathscr{G}$ and $\tilde{\mathscr{G}}$ corresponds merely to the
matrix of the same Hilbert space operator evaluated in two diferent bases, and
the same is true of $\Gamma$ and $\tilde{\Gamma}$.  As such, the matrix 
relation \eqref{eqn:exptildegamma} between $e^{\tilde{\Gamma}}$ and 
$\tilde{\mathscr{G}}$ holds true for $e^{\Gamma}$ and $\mathscr{G}$ 
as well, i.e.~we have
\begin{equation}\label{eqn:expGamma}
e^{\Gamma} = \mathscr{G}(\one - \mathscr{G})^{-1}.
\end{equation}

\subsection{Key Formulas Involving Grassmann Variables}

In the next stage of our derivations, we need to make use of Grassmann 
variables.  These are anticommuting $c$-numbers familiar
in the context of field theory (see for example, Ref.~\onlinecite{negele98}).  
If $\xi_i$ and $\xi_j$ are Grassmann variables, where $i \neq j$, then we have 
$\xi_i\xi_j = -\xi_j\xi_i$ and $\xi_i^2 = 0 = \xi_j^2$.  The purpose of 
introducing these is to define the fermionic coherent states
\begin{equation}
\ket{\boldsymbol{\xi}} = \ket{\xi_1\xi_2\cdots\xi_N} =
\exp\left(-\textstyle\sum_{i=1}^N \xi_i \cxd{i} \right)\ket{0},
\end{equation}
which are eigenstates of the fermion annihilation operators, i.e.~$\cx{i}
\ket{\boldsymbol{\xi}} = -\xi_i\ket{\boldsymbol{\xi}}$.  The value of coherent 
states in general is that one can replace the manipulation of non-commuting 
operators by the manipulation of $c$-number matrix elements.  In the present 
case of fermions, anticommuting operators may be made to commute by
the insertion of Grassmann coefficients. 

There are three key formulas involving Grassmann algebra that we need for the
derivations in this section.  The first involves the matrix element of an
exponentiated bilinear operator $\smash{\exp\left(\sum_{i,j}\Gamma_{ij}\cxd{i}
\cx{j}\right)}$ between fermionic coherent states $\ket{\boldsymbol{\xi}}$ and
$\ket{\boldsymbol{\xi}'}$, given by
\begin{equation}\label{eqn:grassmannmatrixelements}
\braket{\boldsymbol{\xi} | 
\exp\left(\textstyle\sum_{i,j}\Gamma_{ij}\cxd{i}\cx{j}\right) |
\boldsymbol{\xi}'} = 
\exp\left[\textstyle\sum_{i,j}(e^{\Gamma})_{ij}\xi_i^*\xi'_j\right],
\end{equation}
where $e^{\Gamma}$ is the exponential of the matrix $\Gamma$.
The second formula expresses the trace of an operator $A$ as a Grassmann
integral over its coherent state matrix elements as
\begin{equation}\label{eqn:grassmanntrace}
\Tr(A) = \int \prod_i d\xi_i^*\,d\xi_i\,e^{-\sum_i \xi_i^*\xi_i}\,
\braket{-\boldsymbol{\xi}| A | \boldsymbol{\xi} }.
\end{equation}
The third formula that we would need is the Gaussian integral over Grassmann
variables,
\begin{equation}\label{eqn:grassmanngaussianintegral}
\int \prod_i d\xi_i^*\,d\xi_i\,e^{\sum_{j,k}\xi_j^*A_{jk}\xi_k} =
\det A.
\end{equation}

The strategy then would be to evaluate the matrix elements of $\rho_0$ in
\eqref{eqn:rhonoughtgamma} using \eqref{eqn:grassmannmatrixelements}, follow
the prescription in \eqref{eqn:rhoBrhonought} where we trace over the
environmental degrees of freedom using \eqref{eqn:grassmanntrace}, and then
use \eqref{eqn:grassmannmatrixelements} in reverse to recover $\rho_B$ from its
coherent state matrix elements.  Before we do so, let us first tidy up the 
notations by relabelling the coherent states as
\begin{equation}\label{eqn:fermioncoherentstate}
\begin{split}
\ket{\boldsymbol{\xi}\boldsymbol{\eta}} &= 
\ket{\xi_1\cdots\xi_B; \eta_1\cdots\eta_{N-B}} \\
&= \exp\left(-\textstyle\sum_{i=1}^B\xi_i\cxd{i} - 
\textstyle\sum_{j=1}^{N-B}\eta_j\cxd{j}\right)\ket{0},
\end{split}
\end{equation}
where $\boldsymbol{\xi} = \{\xi_1, \dots, \xi_{B}\}$ are Grassmann 
coordinates associated with sites on the block, and $\boldsymbol{\eta} = 
\{\eta_1, \dots, \eta_{N-B}\}$ are Grassmann coordinates associated with sites
in the environment.

\subsection{Matrix Block Form}

Seeing that $\rho_0$ is written in \eqref{eqn:rhonoughtgamma} as the 
exponential of a quadratic form with coefficient matrix $\Gamma$, we make use
of \eqref{eqn:grassmannmatrixelements} to write down its matrix element 
between the fermionic coherent states $\ket{\boldsymbol{\xi}\,
\boldsymbol{\eta}}$ and $\ket{\boldsymbol{\xi}'\,\boldsymbol{\eta}'}$ as
a Gaussian in Grassmann variables:
\begin{equation}
\expn{\boldsymbol{\xi}\,\boldsymbol{\eta}}{\rho_0}{\boldsymbol{\xi}'\,
\boldsymbol{\eta}'} = \mathscr{Q}^{-1}\exp\left[
\begin{pmatrix} \boldsymbol{\xi}^* & \boldsymbol{\eta}^* \end{pmatrix}
e^{\Gamma}
\begin{pmatrix} \boldsymbol{\xi}' \\ \boldsymbol{\eta}' \end{pmatrix}\right].
\end{equation}
Our task now is to derive the matrix elements of $\rho_B$ in the same
Gaussian form, after tracing out the environment.

To find the matrix elements $\expn{\boldsymbol{\xi}}{\rho_B}
{\boldsymbol{\xi}'}$ of the density matrix $\rho_B$ on the block of $B$ sites,
we use \eqref{eqn:grassmanntrace} and perform a partial trace over the
environment to give
\begin{equation}\label{eqn:matrixelements}
\begin{split}
\expn{\boldsymbol{\xi}}{\rho_B}{\boldsymbol{\xi}'} &=
\int d\boldsymbol{\eta}^* d\boldsymbol{\eta}\,
e^{-\boldsymbol{\eta}^*\smallone\boldsymbol{\eta}}\,
\expn{\boldsymbol{\xi}\,{-\boldsymbol{\eta}}}{\rho_0}{\boldsymbol{\xi}'\,
\boldsymbol{\eta}} \\
&= \mathscr{Q}^{-1}\,\int d\boldsymbol{\eta}^* d\boldsymbol{\eta}\,
\times \\
&\qquad\qquad
\exp\left[
\begin{pmatrix} \boldsymbol{\xi}^* & -\boldsymbol{\eta}^* \end{pmatrix}
\begin{pmatrix} \zero & \zero \\ \zero & \one \end{pmatrix}
\begin{pmatrix} \boldsymbol{\xi}' \\ \boldsymbol{\eta} \end{pmatrix} \right]\,
\times \\
&\qquad\qquad
\exp\left[
\begin{pmatrix} \boldsymbol{\xi}^* & -\boldsymbol{\eta}^* \end{pmatrix}
e^{\Gamma}
\begin{pmatrix} \boldsymbol{\xi}' \\ \boldsymbol{\eta} \end{pmatrix}\right] \\
&= \mathscr{Q}^{-1}\,e^{-\boldsymbol{\xi}^*\one\boldsymbol{\xi}'}\,
\int d\boldsymbol{\eta}^* d\boldsymbol{\eta}\, \times \\
&\qquad\qquad
\exp\left[
\begin{pmatrix} \boldsymbol{\xi}^* & -\boldsymbol{\eta}^* \end{pmatrix}
(\one + e^{\Gamma})
\begin{pmatrix} \boldsymbol{\xi}' \\ \boldsymbol{\eta} \end{pmatrix}\right].
\end{split}
\end{equation}
Following this we must express these matrix elements in a form that would 
allow us to trace over the environment.  To do so, let us first write $(\one + 
e^{\Gamma})$ in matrix block form as
\begin{equation}\label{eqn:blockform}
\one + e^{\Gamma} = \begin{bmatrix} A & B \\ B^T & C \end{bmatrix}
\end{equation}
where $A$ is the $B\times B$ \emph{block submatrix}, obtained by restricting 
the indices $i$ and $j$ of $(\one + e^{\Gamma})$ in coordinate space to range 
only over sites on the block, $C$ is the $(N-B)\times(N-B)$ \emph{environment 
submatrix}, obtained by restricting the indices $i$ and $j$ of $(\one + 
e^{\Gamma})$ to range only over sites in the environment, and $B$ is the $B 
\times (N - B)$ \emph{decoherence submatrix} of $(\one + e^{\Gamma})$, 
obtained by restricting the the row index to range only over sites on the 
block and the column index to range only over sites in the environment.

\subsection{Tracing Down \boldmath$\rho_0$}
\label{sect:tracedown}

With \eqref{eqn:matrixelements} and \eqref{eqn:blockform}, the block density
matrix elements can then be written as
\begin{equation}
\begin{split}
\expn{\boldsymbol{\xi}}{\rho_B}{\boldsymbol{\xi}'} &= 
\mathscr{Q}^{-1}\,e^{\boldsymbol{\xi}^*(A - \one)\boldsymbol{\xi}'}\,
\int d\boldsymbol{\eta}^* d\boldsymbol{\eta}\,
e^{\boldsymbol{\xi}^* B \boldsymbol{\eta} - 
\boldsymbol{\eta}^* B^T \boldsymbol{\xi}' -
\boldsymbol{\eta}^* C \boldsymbol{\eta}}.
\end{split}
\end{equation}
Here we have made use of the fact that since the Grassmann variables occur 
quadratically in each term in the exponential, they commute with one another
and we may thus factor the exponential as if it is an exponential of 
$c$-numbers.

By performing a shift of the integration variables $\boldsymbol{\eta}$ and
$\boldsymbol{\eta}^*$, and then evaluating the Grassmann Gaussian integral 
using \eqref{eqn:grassmanngaussianintegral}, we find that
\begin{equation}\label{eqn:blockdmelmts}
\expn{\boldsymbol{\xi}}{\rho_B}{\boldsymbol{\xi}'} = \mathscr{Q}^{-1}
\det C\, e^{\boldsymbol{\xi}^*[A - \one - B C^{-1} B^T]\boldsymbol{\xi}'},
\end{equation}
which parallels Eq.~(14) in Ref.~\onlinecite{chung01}.  From 
\eqref{eqn:blockdmelmts}, we see that the expression for
$\expn{\boldsymbol{\xi}}{\rho_B}{\boldsymbol{\xi}'}$
involves only the Grassman coordinates $\xi_i$ and $\xi'_i$ associated with 
sites on the block.  This is good.  But it also involve the decoherence
submatrix $B$ as well as the environment submatrix $C$, with the latter 
appearing both in the exponential as well as in the normalization constant.  
These matrices have indices that range over sites outside the block, which we
are supposed to have traced out and gotten over with.

Indeed, this must have been successfully done, since $A - \one - B C^{-1} B^T$ 
is a $B \times B$ matrix whose indices range only over sites on the block.
In fact, using \eqref{eqn:blockinversegamma} in Appendix \ref{app:blockinvert},
we can express this matrix entirely in terms of submatrices on the block, and
write \eqref{eqn:blockdmelmts} as
\begin{equation}\label{eqn:almost}
\expn{\boldsymbol{\xi}}{\rho_B}{\boldsymbol{\xi}'} = \mathscr{Q}^{-1}
\det C\,
e^{\boldsymbol{\xi}^*[D^{-1} - \one]\boldsymbol{\xi}'},
\end{equation}
where $D$ is the block submatrix of $(\one + e^{\Gamma})^{-1}$, obtained by
restricting its indices to range only over sites on the block.
That leaves only the $\det C$ in the normalization that we have to deal with.

To express $\mathscr{Q}^{-1}\det C$ in terms of submatrices whose 
indices range only over sites on the block, we make use of the fact that
\begin{equation}
\begin{split}
\Tr(\rho_B) &= 1 = \int d\boldsymbol{\xi}^* d\boldsymbol{\xi}\,
e^{-\boldsymbol{\xi}^*\smallone\boldsymbol{\xi}}
\expn{\boldsymbol{-\xi}}{\rho_B}{\boldsymbol{\xi}} \\
&= \mathscr{Q}^{-1}\det C\,\int d\boldsymbol{\xi}^* d\boldsymbol{\xi}\,
e^{-\boldsymbol{\xi}^* D^{-1}\boldsymbol{\xi}} \\
&= \mathscr{Q}^{-1}\det C\,\det D^{-1},
\end{split}
\end{equation}
which means that
\begin{equation}
\mathscr{Q}^{-1}\det C = \det D.
\end{equation}
With this we have succeeded in writing down a Gaussian form for the coherent 
state matrix elements of $\rho_B$ involving only degrees of freedom on the 
block.  Using the translation machinery provided by 
\eqref{eqn:grassmannmatrixelements}, we then establish the exponentiated form
\begin{equation}
\rho_B = \det D\,\exp\Bigl\{\sum_{i,j}\left[\log(D^{-1} - \one)\right]_{ij}
\cxd{i}\cx{j}\Bigr\}
\end{equation}
of Chung and Peschel.

At this point, let us remark that the above formula for $\rho_B$ is of no
practical use, if to find the matrix $D$, we actually have to evaluate
the matrix $(\one + e^{\Gamma})$, whose indices run over the entire system,
take its inverse $(\one + e^{\Gamma})^{-1}$, and then from this identify the
block submatrix $D$.  This is essentially what was done in 
Ref.~\onlinecite{chung01}, where the matrix parallel to $D^{-1} - \one$ was
computed numerically, for the case of an environment equal in size to the
block.  For our problem, identifying $A - B C^{-1} B^T$ as $D^{-1}$ with
the aid of our analytic relation \eqref{eqn:blockinversegamma} allows us 
to work with arbitrary, even infinite, environment sizes.  

Furthermore, armed with the relationship \eqref{eqn:expGamma} obtained in 
Section~\ref{sect:expformrhonought}, we can find that the normalization and 
matrix of coefficients appearing in \eqref{eqn:almost} in terms of the block 
Green function matrix $G$.  From
\begin{equation}
\one + e^{\Gamma} = \one + \mathscr{G}(\one - \mathscr{G})^{-1} =
(\one - \mathscr{G})^{-1}.
\end{equation}
we see that $D$ is just $(\one - \mathscr{G})$ restricted to the block, 
i.e.~$D = \one - G$, and consequently, $D^{-1} = (\one - G)^{-1}$.  With this, 
the normalization constant for $\rho_B$ can be written as $\det D = 
\det(\one - G)$.  For the matrix of coefficients $(D^{-1} - \one)$ in the
exponential, we see that
\begin{equation}
D^{-1} - \one = (\one - G)^{-1} - \one = G(\one - G)^{-1}.
\end{equation}
With this substitution, the matrix elements of $\rho_B$ now reads as
\begin{equation}
\braket{\boldsymbol{\xi} | \rho_B | \boldsymbol{\xi}' } = 
\det(\one - G)\exp\left[\boldsymbol{\xi}^* G(\one - G)^{-1} 
\boldsymbol{\xi}'\right].
\end{equation}
so that, after using \eqref{eqn:grassmannmatrixelements} in reverse 
translation, we can read off the operator form of $\rho_B$ as
\begin{equation}\label{eqn:operatorform}
\rho_B = \det(\one - G)\,\exp\left\{
\sum_{ij}\left[\log G(\one - G)^{-1}\right]_{ij}
\cxd{i}\cx{j}\right\}.
\end{equation}
In a suitable basis diagonalizing $\log G(\one - G)^{-1}$, this becomes
\begin{equation}\label{eqn:rhoexact}
\rho_B = \det(\one - G)\,\exp\left[-\textstyle\sum_{l}
\varphi_l f_l^{\dagger} f_l\right],
\end{equation}
where the $f_l$'s are linear combinations of $\cx{j}$'s, and $\varphi_l$ is
the associated pseudo-energy (see \eqref{eqn:pseudoenergy} for definition).  
With \eqref{eqn:rhoexact}, we see that to find $\rho_B$, we need only 
calculate the $B\times B$ block Green function matrix $G$ from the ground state 
wavefunction with the aid of operators local to the block, and diagonalize it
to determine $f_l$ and subsequently $\varphi_l$.

To connect this with the results that we obtained in Section~\ref{sect:index},
let us evaluate the matrix elements for the 0- and 1-particle sectors of 
$\rho_B$.  Taylor expanding the exponential in \eqref{eqn:rhoexact} gives us
\begin{equation}
\rho_B = \det(\one - G)\prod_l
\left[\one + (e^{-\varphi_l} - 1)f_l^{\dagger}f_l\right],
\end{equation}
and so we see that the 0-particle sector is given by
\begin{equation}
\rho_{B,0} = {}_B\!\braket{0 | \rho_B | 0 }_B = \det(\one - G),
\end{equation}
while in the basis diagonalizing $\rho_B$, the matrix elements in the 
1-particle sector are given by
\begin{equation}\label{eqn:onepartsect}
\begin{split}
{}_B\!\braket{0|f_l \rho_{B,1} f_l^{\dagger} |0}_B &= \det(\one - G)
\biggl[
{}_B\!\braket{0|f_l \one f_l^{\dagger}|0}_B + \\
&\quad\ \sum_{l'} (e^{-\varphi_{l'}} - 1)
{}_B\!\braket{0|f_l f_{l'}^{\dagger} f_{l'} f_l^{\dagger}|0}_B \biggr] \\
&= \det(\one - G)\,e^{-\varphi_l} \\
&= \det(\one - G)\,\left[G(\one - G)^{-1}\right]_{ll}.
\end{split}
\end{equation}
This completes the proof of our conjecture at the end of 
Section~\ref{sect:dmblock} that as a matrix, $\rho_{B,1}$ is related to 
$G$ by \eqref{eqn:closedform}.

\subsection{The Pseudo-Energies $\boldsymbol{\varphi_l}$}

With the closed-form formula \eqref{eqn:rhoexact} for $\rho_B$ at hand, we are 
now ready to understand its structure and spectra.  To begin with, we find 
that the exponential form 
\begin{equation}
\begin{split}
\rho_B &= \det(\one - G)\exp\left[\sum_{ij}
\left(\log G(\one - G)^{-1}\right)_{ij}\cxd{i}\cx{j}\right] \\
&= \det(\one - G)\exp\left[-\sum_{ij}\Phi_{ij}\cxd{i}\cx{j}\right],
\end{split}
\end{equation}
where we define the matrix $\Phi$ to be
\begin{equation}\label{eqn:Phi}
\Phi = -\log G(\one - G)^{-1} = -\log G + \log(\one - G),
\end{equation}
implies that the weights and eigenvectors of the $(F > 1)$-particle sectors of 
$\rho_B$ are determined completely by those in the 0- and 1-particle sectors.  
Defining the set of \emph{pseudo-energies}
\begin{equation}\label{eqn:pseudoenergy}
\varphi_l = -(\log G(\one - G)^{-1})_{ll},
\end{equation}
for $l = 1, \dots, B$, which are the eigenvalues of $\Phi$, and $\varphi_0 = 
-\log\det(\one - G)$, we find that the weights of the 1-particle block can 
be written as
\begin{equation}\label{eqn:wandphi}
w_l = \exp[-(\varphi_0 + \varphi_l)],
\end{equation}
and $\rho_B$ can be written in the form
\begin{equation}
\rho_B = e^{-\varphi_0}\exp\left[-\sum_l \varphi_l f_l^{\dagger} f_l\right] =
e^{-\varphi_0}\exp[-\tilde{H}].
\end{equation}
Inspired by the resemblance of the form of $\rho_B$ to the density matrix of a 
quantum canonical ensemble, we call $\tilde{H}$ the \emph{pseudo-Hamiltonian}. 

\subsection{Particle-Hole Symmetry at Half-Filling}
\label{sect:partholesym}

To complete our understanding of the structure and spectrum of $\rho_B$, we
want to know how symmetries of the original problem are built into $\rho_B$.
In particular, we will consider particle-hole symmetry on a bipartite lattice,
on which we define a `charge-conjugation' operator $\mathscr{C}$, with 
$\mathscr{C}^2 = \one$.\footnote{One possible form for the charge-conjugation 
operator is $\mathscr{C} = \prod_j \left[i^{j+1} \protect\cxd{j} + 
(-i)^{j+1} \protect\cx{j}\right]$, where the product runs over all lattice 
sites.} The action of $\mathscr{C}$ on the coordinate space
fermion operators can be defined to be
\begin{equation}\label{eqn:action}
\mathscr{C}\cx{i}\mathscr{C} = -(-1)^i\cxd{i},\quad
\mathscr{C}\cxd{i}\mathscr{C} = -(-1)^i\cx{i},
\end{equation}
where $(-1)^i$ is defined to be $+1$ ($-1$) whenever the site $i$ belongs to
the even (odd) sublattice.  In a $d$-dimensional hypercubic lattice, where the 
site index is $i = \{i_1, i_2, \dots, i_d\}$, the factor $(-1)^i$ is rightfully 
given by $(-1)^i = (-1)^{i_1 + i_2 + \cdots + i_d}$.

There are two conditions, one on the dispersion relation $\epsilon_k$, and the 
other on the chemical potential $\mu$, implied by particle-hole symmetry.  To
derive the first condition on the dispersion relation, we note from
\eqref{eqn:action} that in momentum space --- when the lattice is a Bravais 
lattice --- that
\begin{equation}\label{eqn:kaction}
\mathscr{C} \ck{k} \mathscr{C} = -\ckd{-k-Q}, \qquad
\mathscr{C} \ckd{k} \mathscr{C} = -\ck{-k+Q},
\end{equation}
where the wavevector $Q$ is defined by $e^{i Q\cdot r_i} = 
(-1)^i$.\footnote{If the bipartite lattice is not a Bravais lattice, then
whereever the wavevector $k$ appears as an index, it must be replaced by
the combination of $k$ and a band index.  All of the results --- in particular
those of Section \ref{sect:partholesym} --- still go through in this
generalized case, provided that all lattice sites are symmetry equivalent.}
We can then check, with \eqref{eqn:hamfourier} and \eqref{eqn:kaction}, that
\begin{equation}
\mathscr{C} H \mathscr{C} = \sum_k \epsilon_k \ck{-k+Q}\ckd{-k-Q} 
= \sum_{k'} \epsilon_{-k'-Q}\ck{k'+2Q}\ckd{k'}.
\end{equation}
Now, from the definition of $Q$, it is clear that
\begin{equation}
\begin{split}
\ck{k'+2Q} &= N^{-1/2}\sum_j e^{i(k'+2Q)\cdot r_j}\,\cx{j} \\
&= N^{-1/2} \sum_j e^{ik'\cdot r_j}[(-1)^j]^2\,\cx{j} = \ck{k'},
\end{split}
\end{equation}
and thus (dropping the prime on the dummy wavevector $k'$ that is summed over)
\begin{equation}\label{eqn:CHC}
\mathscr{C} H \mathscr{C} = \sum_k \epsilon_{-k-Q}\ck{k}\ckd{k} =
\sum_k \epsilon_{-k-Q} + \sum_k -\epsilon_{-k-Q}\ckd{k}\ck{k}.
\end{equation}
For time-reversal invariant systems, $\epsilon_{-k} = \epsilon_k$.  Also,
for our choice of Hamiltonian, $\sum_k \epsilon_{-k-Q} = 
\sum_{k'} \epsilon_{k'} = \Tr H = 0$. Thus \eqref{eqn:CHC} simplifies to
\begin{equation}\label{eqn:CHCk}
\mathscr{C} H \mathscr{C} = \sum_k -\epsilon_{k+Q}\ckd{k}\ck{k}.
\end{equation}
Since it is clear from \eqref{eqn:hamiltonian} and \eqref{eqn:action} that
$\mathscr{C} H \mathscr{C} = H$, \eqref{eqn:CHCk} tells us that the dispersion 
relation associated with the particle-hole symmetric Hamiltonian $H$ must 
satisfy the condition
\begin{equation}\label{eqn:ekparthole}
\epsilon_{k+Q} = -\epsilon_k.
\end{equation}

Next, to understand how the second condition on the chemical potential comes 
about, let us note the trivial fact that, since $\rho_B$ is a reduced density 
matrix of $\rho_0$, for there to be any sense in talking about the 
manifestation of particle-hole symmetry in $\rho_B$, $\rho_0$ must first be 
particle-hole symmetric, i.e.~$\mathscr{C} \rho_0 \mathscr{C} = \rho_0$.  When 
$\rho_0$ is the density matrix of the ground state at $T = 0$, then it is 
particle-hole symmetric whenever the ground state $\ket{\Psi_F}$ is.  For
$\ket{\Psi_F}$ to be particle-hole symmetric, it must have the transformation
property
\begin{equation}\label{eqn:halffilledgroundstate}
\mathscr{C}\ket{\Psi_F} = \eta_{\mathscr{C}}\ket{\Psi_F},
\end{equation}
where $\eta_{\mathscr{C}} = \pm 1$ is a phase factor associated with 
$\mathscr{C}$.  We know that this is satisfied only by the half-filled 
ground state.  At finite temperature, when $\rho_0$ is 
taken from the grand canonical ensemble and has the form given in 
\eqref{eqn:rho0}, what, if any, extra conditions must be satisfied in order 
for $\rho_0$ to be particle-hole symmetric?

Indeed, there appears to be cause for concern: unlike $H$, which is invariant
under `charge-conjugation', the fermion number operator $F$ transforms under
$\mathscr{C}$ as
\begin{equation}
\mathscr{C} F \mathscr{C} = \sum_i \mathscr{C} \cxd{i}\cx{i} \mathscr{C} =
\sum_i \cx{i}\cxd{i} = N - F,
\end{equation}
and so for $\rho_0$ to be particle-hole symmetric, i.e.
\begin{equation}
\mathscr{C} \rho_0 \mathscr{C} = \mathscr{Q}^{-1}\exp \beta [
H - \mu (N - F)] = \rho_0,
\end{equation}
we must have $\mu = 0$.  For a dispersion relation satisfying
\eqref{eqn:ekparthole}, $\mu = 0$ corresponds to precisely the situation of
half-filling.  At least for the grand canonical ensemble, there appears to 
be no other conditions necessary for $\rho_0$ to be particle-hole symmetric. 

With these conditions in mind, we may now proceed to investigate how 
particle-hole symmetry shows up in the pseudo-energy spectrum (and hence the 
spectrum of the block density matrix $\rho_B$).  But first, we must understand 
how the symmetry is manifested in the Green function matrix
$\mathscr{G}$, and its restriction to the block, $G$.  Knowing from our
arguments above that $\mu = 0$, we see that the matrix elements of 
$\tilde{\mathscr{G}}$ in momentum space simplifies to
\begin{equation}
\tilde{\mathscr{G}}_{kk} = \frac{1}{\exp\beta\epsilon_k + 1}.
\end{equation}
Furthermore, using \eqref{eqn:ekparthole}, we can relate 
$\tilde{\mathscr{G}}_{k+Q, k+Q}$ to $\tilde{\mathscr{G}}_{kk}$ by
\begin{equation}
\tilde{\mathscr{G}}_{k+Q,k+Q} = \frac{1}{\exp\beta\epsilon_{k+Q} + 1} =
\frac{1}{\exp(-\beta\epsilon_k) + 1} = 1 - \tilde{\mathscr{G}}_{kk}.
\end{equation}
This gives rise to the condition
\begin{equation}\label{eqn:JGJ}
\mathscr{G}_{ij} = \delta_{ij} - (-1)^{(i-j)}\mathscr{G}_{ij}
\end{equation}
that must be satisfied by the matrix elements of $\mathscr{G}$ in coordinate
space.

This same result can be derived more transparently for the special case of
$T = 0$: using the fact that $\mathscr{C}^2 = \one$, $\eta_{\mathscr{C}}^2 = 
1$, as well as \eqref{eqn:action} and \eqref{eqn:halffilledgroundstate}, 
we find that
\begin{equation}\label{eqn:JGJb}
\begin{split}
\mathscr{G}_{ij} &= \braket{\Psi_F|\cxd{i}\cx{j}|\Psi_F} \\
&= \braket{\Psi_F|\mathscr{C}(\mathscr{C}\cxd{i}\mathscr{C})
(\mathscr{C}\cx{j}\mathscr{C})\mathscr{C}|\Psi_F} \\
&= (-1)^{i+j}\braket{\Psi_F|\cx{i}\cxd{j}|\Psi_F} \\
&= (-1)^{i+j}\delta_{ij} - (-1)^{i+j}\braket{\Psi_F|\cxd{j}\cx{i}|\Psi_F} \\
&= \delta_{ij} - (-1)^{i+j} \mathscr{G}_{ij},
\end{split}
\end{equation}
where we have made use of the fact that $\mathscr{G}$ is symmetric, 
i.e.~$\mathscr{G}_{ji} = \mathscr{G}_{ij}$.

Since \eqref{eqn:JGJb} is a condition satisfied by the matrix elements of
$\mathscr{G}$ individually, it holds just as well to those restricted to the
block, i.e.~$G_{ij}$.  In particular, we note that \eqref{eqn:JGJb} can
actually be written as a matrix equation, which when restricted to the block
reads as
\begin{equation}\label{eqn:halffilledG}
G = \one - JGJ,
\end{equation}
where $J = \diag(e^{iQ\cdot r_i}) = \diag(1,-1,1,-1,\dots)$ in coordinate 
space, and $J^2 = \one$.

To appreciate the implications of \eqref{eqn:halffilledG}, let us consider an
eigenvector $\ket{\lambda_l}$ of $G$ correspond to the
eigenvalue $\lambda_l$, such that
\begin{equation}
G\ket{\lambda_l} = \lambda_l\ket{\lambda_l}.
\end{equation}
By \eqref{eqn:Phi}, this is also the eigenvector of $\rho_B$, with 
corresponding pseudo-energy
\begin{equation}
\varphi_l = -\log\lambda_l + \log(1 - \lambda_l).
\end{equation}
Using \eqref{eqn:halffilledG}, we find that
\begin{equation}
\begin{split}
G J\ket{\lambda_l} &= (\one - J G J) J\ket{\lambda_l} \\
&= J\ket{\lambda_l} - J G J^2 \ket{\lambda_l} \\
&= J\ket{\lambda_l} - J G \ket{\lambda_l} \\
&= J\ket{\lambda_l} - \lambda_l J \ket{\lambda_l} \\
&= (1 - \lambda_l) J \ket{\lambda_l},
\end{split}
\end{equation}
i.e.~the state $\ket{\lambda_{l'}} \equiv J\ket{\lambda_l}$ generated by
particle-hole symmetry from $\ket{\lambda_l}$ is also an eigenvector of
$G$, with eigenvalue $\lambda_{l'} = (1 - \lambda_l)$.  The pseudo-energy
$\varphi_{l'}$ associated with $\ket{\lambda_{l'}}$ is then
\begin{equation}
\varphi_{l'} = -\log\lambda_{l'} + \log(1 - \lambda_{l'}) = -\varphi_l.
\end{equation}
It is interesting to note how the weights $w_{B,1,l}$, being exponentials of
the pseudo-energies $\varphi_l$, hide this particular aspect of 
particle-hole symmetry.

\section{The $\boldsymbol{(F > 1)}$-Particle Sectors}
\label{sect:checks}

Up to this point, our discussions have been for arbitrary dimensions.  To
demonstrate how the $\smash{(F > 1)}$-particle sectors can be constructed 
from the 0- and 1-particle sectors, we specialize to the 1-dimensional case, 
wherein the Fermi sea is
\begin{equation}
\ket{\Psi_F} = \prod_{|k| = 0}^{|k| = \nbar(\pi/a)} \ckd{k}\ket{0},
\end{equation}
where $a$ is the lattice constant and $\nbar$ is the filling fraction.  The
2-point functions can be computed explicitly as
\begin{equation}
G_{\bar ij}  = \frac{\sin\pi\nbar|i-j|}{\pi|i-j|}.
\end{equation}
We now illustrate how to construct the weights and eigenvectors of the 
$(F > 1)$-particle sectors of $\rho_B$ starting from $\varphi_0$, $\varphi_l$ 
and $f_l$, using the example of $B = 3$ at half-filling.

At half-filling, $\nbar = \frac{1}{2}$, the 2-point functions $G_{\bar ij}$
take on particularly simple values
\begin{equation}
G_{\bar 11} = G_{\bar 22} = G_{\bar 33} = \tfrac{1}{2},
\quad
G_{\bar 12} = G_{\bar 23} = \tfrac{1}{\pi},
\quad
G_{\bar 13} = 0,
\end{equation}
with which we find, using the machinery developed in 
Section~\ref{sect:freespinlessfermions}, the 0-particle and 1-particle sectors 
of $\rho_3$ to be
\begin{equation}
\begin{split}
\rho_{3,0} &= \braket{000|\rho_3|000} = \tfrac{1}{8} - \tfrac{1}{\pi^2}, \\
\rho_{3,1} &= \begin{bmatrix}
\braket{100|\rho_3|100} & \braket{100|\rho_3|010} & 
\braket{100|\rho_3|001} \\[1.2ex]
\braket{010|\rho_3|100} & \braket{010|\rho_3|010} & 
\braket{010|\rho_3|001} \\[1.2ex]
\braket{001|\rho_3|100} & \braket{001|\rho_3|010} & \braket{001|\rho_3|001}
\end{bmatrix} \\
&= \begin{bmatrix}
\frac{1}{8} & \frac{1}{2\pi} & \frac{1}{\pi^2} \\[1.2ex]
\frac{1}{2\pi} & \frac{1}{8} + \frac{1}{\pi^2} & \frac{1}{2\pi} \\[1.2ex]
\frac{1}{\pi^2} & \frac{1}{2\pi} & \frac{1}{8} \end{bmatrix}.
\end{split}
\end{equation}

Diagonalizing these, we find
\begin{subequations}
\begin{align}
w_{3,0,1} &= \left(\tfrac{1}{\sqrt{8}} - \tfrac{1}{\pi}\right)
\left(\tfrac{1}{\sqrt{8}} + \tfrac{1}{\pi}\right), \\
w_{3,1,1} &= \left(\tfrac{1}{\sqrt{8}} + \tfrac{1}{\pi}\right)^2, \notag \\
w_{3,1,2} &= \left(\tfrac{1}{\sqrt{8}} - \tfrac{1}{\pi}\right)
\left(\tfrac{1}{\sqrt{8}} + \tfrac{1}{\pi}\right), \quad
\smash{\left.\begin{matrix} \\ \\ \\ \\ \\ \\ \end{matrix}\right\}} \\
w_{3,1,3} &= \left(\tfrac{1}{\sqrt{8}} - \tfrac{1}{\pi}\right)^2, \notag
\end{align}
\end{subequations}
and thus
\begin{subequations}
\begin{align}
\varphi_0 &= -\log\left(\tfrac{1}{\sqrt{8}} - \tfrac{1}{\pi}\right)
\left(\tfrac{1}{\sqrt{8}} + \tfrac{1}{\pi}\right) = +3.74317\dots, \\
\varphi_1 &= -\log\dfrac{\tfrac{1}{\sqrt{8}} + \tfrac{1}{\pi}}
{\tfrac{1}{\sqrt{8}} - \tfrac{1}{\pi}} = -2.94777\dots, \notag \\
\varphi_2 &= -\log 1 = 0, \qquad\qquad\qquad\qquad\quad
\smash{\left.\begin{matrix} \\ \\ \\ \\ \\ \\ \\ \\ \end{matrix}\right\}}\\
\varphi_3 &= -\log\dfrac{\tfrac{1}{\sqrt{8}} - \tfrac{1}{\pi}}
{\tfrac{1}{\sqrt{8}} + \tfrac{1}{\pi}} = +2.94777\dots, \notag
\end{align}
\end{subequations}
Since $\varphi_1 = -\varphi_3$, we call these a \emph{particle-hole conjugate
pair} of pseudo-energies, and say that $\varphi_3$ is the particle-hole
conjugate of $\varphi_1$.
The eigenvectors of the 1-particle sector are
\begin{subequations}
\begin{align}
f_1^{\dagger} &= \tfrac{1}{2}\cxd{1} + \tfrac{1}{\sqrt{2}}\cxd{2} +
\tfrac{1}{2}\cxd{3}, \\
f_2^{\dagger} &= \tfrac{1}{\sqrt{2}}\cxd{1} - \tfrac{1}{\sqrt{2}}\cxd{3}, \\
f_3^{\dagger} &= \tfrac{1}{2}\cxd{1} - \tfrac{1}{\sqrt{2}}\cxd{2} +
\tfrac{1}{2}\cxd{3},
\end{align}
\end{subequations}
corresponding to $\varphi_1$, $\varphi_2$ and $\varphi_3$ respectively.

We can easily check that the $f_l^{\dagger}$'s obey the same anticommutator
relation as the $\cxd{i}$'s, i.e.~they obey Pauli's Exclusion Principle, and
so the eigenvectors of the 2-particle sector of $\rho_3$ are created by
\begin{subequations}
\begin{align}
f_1^{\dagger}f_2^{\dagger} &= -\tfrac{1}{2}\cxd{2}\cxd{3} - 
\tfrac{1}{\sqrt{2}}\cxd{1}\cxd{3} - \tfrac{1}{2}\cxd{1}\cxd{2}, \\
f_1^{\dagger}f_3^{\dagger} &= \tfrac{1}{\sqrt{2}}\cxd{2}\cxd{3} -
\tfrac{1}{\sqrt{2}}\cxd{1}\cxd{2}, \\
f_2^{\dagger}f_3^{\dagger} &= -\tfrac{1}{2}\cxd{2}\cxd{3} + \tfrac{1}{\sqrt{2}}
\cxd{1}\cxd{3} - \tfrac{1}{2}\cxd{1}\cxd{2},
\end{align}
\end{subequations}
with associated pseudo-energies $\varphi_1 + \varphi_2 = \varphi_1$,
$\varphi_1 + \varphi_3 = 0 = \varphi_2$ and $\varphi_2 + \varphi_3 = \varphi_3$
respectively.  Here we see that because of the particle-hole symmetry in the
ground state wavefunction, the pseudo-energies of the 2-particle sector
are identical to those in the 1-particle sector, which implies that the
density matrix weights of the 2-particle sector are identical to those in
the 1-particle sector.

For the 3-particle sector, we find that the eigenvector is created by the
operator $f_1^{\dagger}f_2^{\dagger}f_3^{\dagger} = \cxd{1}\cxd{2}\cxd{3}$,
associated with pseudo-energy $\varphi_1 + \varphi_2 + \varphi_3 = 0$, and
hence $w_{3,3,1} = e^{-\varphi_0} = w_{3,0,1}$.
This method of generating all $(F > 1)$-particle sectors, starting from the 0- 
and 1-particle sectors, for larger block sizes at various filling fractions 
$\nbar$ was verified numerically.

Another manifestation of particle-hole symmetry is a queer degeneracy between
weights in the $F$- and $(F+2)$-particle sectors.  This we understand as
follows: if $\varphi_{l_1} + \cdots + \varphi_{l_F}$ is a weight
in the $F$-particle sector, then in general we can find weights 
$(\varphi_{l_1} + \cdots + \varphi_{l_F}) + \varphi_{l_{F+1}} + 
\varphi_{l_{F+2}} = (\varphi_{l_1} + \cdots + \varphi_{l_F})$ in the 
$(F+2)$-particle sector, where $\varphi_{l_{F+1}}$ and $\varphi_{l_{F+2}}$ are 
particle-hole conjugates of each other.

In fact, from the construction outlined above, we also know the pattern of 
degeneracy.  For example, for $B = 6$ at half-filling, the pseudo-energies are 
of the form $-\varphi_a$, $-\varphi_b$, $-\varphi_c$, $\varphi_c$, $\varphi_b$ 
and $\varphi_a$, corresponding to the eigenstates created by 
$f_{-a}^{\dagger}$, $f_{-b}^{\dagger}$, $f_{-c}^{\dagger}$, $f_c^{\dagger}$, 
$f_b^{\dagger}$ and $f_a^{\dagger}$ respectively, where $\varphi_a > \varphi_b 
> \varphi_c$.  We then see in the 3-particle sector that $f_b^{\dagger}
f_{-b}^{\dagger}f_{-a}^{\dagger}\ket{0}$ and $f_c^{\dagger}f_{-c}^{\dagger}
f_{-a}^{\dagger}\ket{0}$ have the same pseudo-energy of $\varphi_b - \varphi_b 
- \varphi_a = \varphi_a = \varphi_c - \varphi_c + \varphi_a$, and are thus 
degenerate, whereas $f_c^{\dagger}f_{-b}^{\dagger}f_{-a}^{\dagger}\ket{0}$ is 
nondegenerate with pseudo-energy $\varphi_c - \varphi_b - \varphi_a$.

\section{Conclusions}

To summarize, in this paper we showed that elements of the block density
matrix, $(\rho_B)_{b'b}$, can be calculated as the expectation 
$\mean{K_b^{\dagger}K_{b'}}$ of a product of referencing operators $K_b$
and $K_{b'}$, which are themselves formed out of fermion operators $\cx{j}$
and $\cxd{j}$ local to the block.  By inspecting the matrix elements
$(\rho_{B,1})_{ij}$ and $G_{\bar ij}$ of the 1-particle sector of $\rho_B$ and 
the block Green function matrix $G$ respectively for block sizes up to $B = 5$,
we are led to a conjecture of the closed-form relation \eqref{eqn:closedform}
between $\rho_{B,1}$ and $G$.

Adapting the technique that Chung and Peschel used to calculate the half-chain
density matrix of a chain of spinless Bogoliubov fermions, we find that we
can not only prove this conjecture, but also derive a closed-form relation 
\eqref{eqn:operatorform} between the entire block density matrix $\rho_B$ and 
$G$, which can also be written in \eqref{eqn:rhoexact} as the exponential of
a pseudo-Hamiltonian $\tilde{H}$.  The spectrum of $\tilde{H}$ is generated
by the independent fermion operators $f_l^{\dagger}$, which also generate the
eigenvectors of $G$, and hence can be determined by diagonalizing $G$.  It is
amusing to numerically compute the pseudo-Hamiltonian in real space.  For
$\nbar \neq \frac{1}{2}$, the generic form of $\tilde{H}$ (a bilinear in
$\{\cxd{j}\}$ and $\{\cx{i}\}$) admits hopping to all other sites on the
block, as well as an on-site potential.  The symmetry at half-filling
ensures that the on-site potential is zero and hopping only connects to the
other sublattice.

We identify three important implications of \eqref{eqn:rhoexact} in formulating
truncation schemes based on $\rho_B$, for the purpose of performing an RG
analysis.  Firstly, we note that the spectrum of $\rho_B$ is completely
determined by the block Green function matrix $G$.  It suffice therefore to
calculate the eigenvectors and eigenvalues in its 0- and 1-particle sectors.  
The eigenvectors and eigenvalues of the $(F > 1)$-particle sectors of $\rho_B$,
if needed, can be systematically generated from the fermion operators 
$f_l^{\dagger}$ and their pseudo-energies $\varphi_l$, as illustrated in 
Section~\ref{sect:checks}.  This fact is evident in Ref.~\onlinecite{chung01},
but its significance was not emphasized.  If one is studying the density 
matrix of a noninteracting toy model (as in this paper), we have an enormous
saving in terms of computational time: instead of diagonalizing the entire 
$\rho_B$, which is of rank $O(e^B)$, we can diagonalize just the 1-particle 
sector $\rho_{B,1}$, which is of rank $O(B)$.  Possible objects of such a
study could be: (i) the distribution of eigenvalues;\cite{cheong} (ii) errors
in the dispersion relation due to truncation;\cite{cheong} and (iii) comparing 
the product basis of two blocks of length $B$ with the basis of one block of 
length $2B$, to weigh the effects of the correlations respectively neglected 
or included.

Secondly, it is highly desirable in RG calculations to ensure that the 
truncation scheme preserves the symmetries of the target state.  Using the
specific example of particle-hole symmetry, we saw in Section~\ref{sect:checks} 
that $\rho_{B,F}$ and $\rho_{B,B-F}$ have the same set of weights, and the
eigenvectors of $\rho_{B,F}$ are related, up to a phase, to the 
eigenvectors of $\rho_{B,B-F}$ acted upon by $\mathscr{C}$.  Naively, we 
might expect that to preserve particle-hole symmetry, all we have to do is
to keep $\mathscr{C}\ket{w_{B,F,k}}$ in $\rho_{B,B-F}$ if $\ket{w_{B,F,k}}$
in $\rho_{B,F}$ is kept.  However, there is more to particle-hole
symmetry.  Under the action of $\mathscr{C}$, the half-filled ground state 
$\ket{\Psi_F}$ goes (up to a phase) back to itself.  Within the block, this
global symmetry transformation brings the mixed state of the block back to the
same mixed state.  Because the mixed state of the block does not have a
definite particle number, particle-hole symmetry is not merely a relation
between $\rho_{B,F}$ and $\rho_{B,B-F}$.  Rather, particle-hole symmetry
imposes strict conditions on the spectra of $\rho_{B,F}$ and $\rho_{B,F'}$,
for $F, F' = 0, \dots, B$.  In fact, in Section~\ref{sect:partholesym}, we
elaborated on the condition that particle-hole symmetry imposes on the 
1-particle sector.  This condition is most intuitive when written in terms of 
the eigenvalues $\lambda_l$ of $G$ or the pseudo-energies $\varphi_l$, but not 
immediately apparent if we just stare at the 1-particle density matrix
weights $w_{B,1,l}$.  It is therefore dangerous to base symmetry-preserving 
truncation schemes on $\rho_B$ and its eigenvalues alone.

This brings us to the last of the implications that we wish to highlight. 
While a toy noninteracting model is studied in this paper, our ultimate 
goal is to address interacting systems, particularly Fermi liquids.  Since 
these (in their low-energy limit) have the same eigenstate
structure as a noninteracting Fermi sea (after a unitary transformation), 
their density matrices also should have the same structure as a noninteracting
system.  The explicit form of the many-body density matrix, as exhibited in
Section \ref{sect:exact} of this paper, hints at the proper design of
truncation schemes.  Rather than independently truncating in each $F$-particle
sector, we should define the truncated states using a set of `creation
operators' which satisfy the usual anticommutation relations, and quite likely
these are closely related to the approximate quasiparticle creation operators,
which should be constructed as a product of the renormalization scheme.
We will have more discussions on the implications of such a truncation
scheme based on picking out a set of appropriate `creation operators' for
the numerical study of interacting systems, the role of dimensionality, and
comparisons with the conventional DMRG, in a second paper.\cite{cheong}

Based on our observations on the pattern of degeneracies within and between 
the $F$-particle sectors of $\rho_B$ in Section~\ref{sect:checks}, we realize 
that if the truncation is carried out naively, there is a very real danger of 
ending up with an inconsistent scheme of truncation.  This problem occurs 
quite generally, at various filling fractions and block sizes, but can be most 
clearly illustrated using our example of $B = 6$ at half-filling.  For example, 
let us say that as the result of a naive truncation, the states 
$f_{-a}^{\dagger}\ket{0}$, 
$f_{-b}^{\dagger}\ket{0}$, $f_{-c}^{\dagger}\ket{0}$ and $f_c^{\dagger}\ket{0}$ 
in the 1-particle sector are kept.  Examining the 2-particle sector, we find
the states $f_{-a}^{\dagger}f_a^{\dagger}\ket{0}$ and $f_{-c}^{\dagger}
f_c^{\dagger}\ket{0}$, which are degenerate in their pseudo-energies.
We can build up the latter, but not the former, using the 1-particle operators 
kept, and so we should keep the latter but not the former.  If we truncate
the 2-particle sector naively, then based on the density matrix weights alone
we would be probably end up keeping or throwing out both of these 2-particle 
states.

In fact, the situation for naive truncation is worse, since the state 
$f_{-a}^{\dagger}f_b^{\dagger}\ket{0}$ has lower pseudo-energy than 
$f_{-c}^{\dagger}f_c^{\dagger}\ket{0}$ and will be kept instead.  We see 
therefore that naive truncation is likely to led to inconsistencies: some
many-particle states built up from the 1-particle states kept get thrown out, 
while other many-particle states that cannot be built up from the set of 
1-particle states kept end up being retained.  Hence, we find that as far as 
particle-conserving models are concerned, for any truncation scheme to be 
consistent, the truncation must be carried out on the 1-particle sector of 
$\rho_B$ alone.

Finally, let us remark that everything done in this paper can be trivially 
extended to the case of spinfull fermions, so long as they are noninteracting. 
Every object in our calculations, in particular the Fermi sea wavefunction and 
the block density matrix, will merely in the spinfull case be replaced by the 
direct product of two such objects with spin-up and spin-down flavors. 

\begin{acknowledgments}
SAC would like to thank Mr.~Hway Kiong Lim for extending many helpful 
suggestions.  This research is supported by NSF grant DMR-9981744, and
made use of the computing facility of the Cornell Center for Materials 
Research (CCMR) with support from the National Science Foundation Materials 
Research Science and Engineering Centers (MRSEC) program (DMR-0079992).
\end{acknowledgments}

\begin{appendix}

\section{Automating the Numerical Computation of $\boldsymbol{\rho_B}$}
\label{app:automatic}

As we saw in Sec.~\ref{sect:dmblock}, each of the $2^B$ basis configurations 
of the block corresponds to an operator $K_b$, so that $(\rho_B)_{b'b} = 
\mean{ K_{b}^{\dagger}K_{b'} }$.
Therefore, to obtain all matrix elements of $\rho_B$ it sufficed
to automate the calculation of expectations of an arbitrary
string of creation/annihilation operators (each operator
acting on one site).

First, this formal string of operators must be simplified.  Through a 
systematic set of anticommutations, it is brought to a canonical form, such 
that (a) it is normal-ordered, with one substring of all creation operators 
followed by one substring of all annihilation operators; and (b) within each 
substring the operators are ordered by the site.  Of course, each site
can appear at most once in each substring (otherwise it reduces to zero.)  A 
complication of this step is that the result is generally a sum of many terms 
in the canonical form, since every rearrangement of the form $\cx{1}\cxd{1}
\to \one - \cxd{1} \cx{1}$ produces two terms from one.

Next, we note that within this sum, only terms containing a balanced number,
say $n$, each of creation and annihilation operators will contribute to the
expectation.  By the Wick theorem, such $2n$-point functions $G_{\bar i_1
\cdots \bar i_n j_1 \cdots j_n}$ reduces to the determinant of an $n\times n$ 
matrix, as shown in \eqref{eqn:twonptfunctions}. 

For models in which fermion number $F$ is conserved, we can further separate 
$\rho_B$ into the various $F$-particle sectors $\rho_{B,F}$ before 
diagonalization.  This is particularly importantly at half-filling, for there 
exists generic degeneracies between states in different sectors (see
Section~\ref{sect:checks}), and there is a danger that a naive diagonalization
of the whole matrix $\rho_B$ will produce eigenstates with mixed particle 
number.

The limiting consideration for the whole calculation is the 
diagonalization time, which is determined by the condition number of $\rho_B$,
rather than array storage.  In general, the condition number, which is the
ratio of the largest weight to the smallest west, grows exponentially with
system size.\cite{cheong}

\section{Block Inversion Formula}
\label{app:blockinvert}

Consider a square $N\times N$ symmetric matrix $\mathscr{M}$ written in 
matrix block form as
\begin{equation}
\mathscr{M} = \begin{bmatrix} A & B \\ B^T & C \end{bmatrix},
\end{equation}
where $A$ is a square $N_1\times N_1$ symmetric matrix, $B$ is a $N_1\times
N_2$ non-square matrix and $C$ is a square $N_2\times N_2$ symmetric matrix.
Here $N_1 + N_2 = N$.

If we write the inverse matrix $\mathscr{M}^{-1}$ also in the matrix block form
\begin{equation}
\mathscr{M}^{-1} = \begin{bmatrix} D & E \\ E^T & F \end{bmatrix},
\end{equation}
where $D$ is a square $N_1\times N_1$ symmetric matrix, $E$ is a $N_1\times
N_2$ non-square matrix and $F$ is a square $N_2\times N_2$ symmetric matrix,
how are $D$, $E$ and $F$ related to the matrix blocks $A$, $B$ and $C$ in 
$\mathscr{M}$?

Using the fact that $\mathscr{M}\mathscr{M}^{-1} = \one$, and thus
\begin{equation}
\begin{bmatrix} A & B \\ B^T & C \end{bmatrix}
\begin{bmatrix} D & E \\ E^T & F \end{bmatrix} =
\begin{bmatrix} \one_{N_1\times N_1} & \zero_{N_1\times N_2} \\ 
\zero_{N_2\times N_1} & \one_{N_2\times N_2} \end{bmatrix},
\end{equation}
(where the subscripts, which will henceforth be dropped for notational clarity,
following the $\one$'s and $\zero$'s indicate the shape and size of the 
matrices) we find the following relations between the matrix blocks of 
$\mathscr{M}$ and $\mathscr{M}^{-1}$:
\begin{subequations}
\begin{align}
AD + BE^T &= \one, \\
AE + BF &= \zero, \\
B^TD + CE^T &= \zero, \\
B^TE + CF &= \one.
\end{align}
\end{subequations}
Solving for $D$, $E$ and $F$ in terms of $A$, $B$ and $C$, we find that
\begin{subequations}\label{eqn:blockinversegamma}
\begin{align}
D &= \left[A - BC^{-1}B^T\right]^{-1}, \\
E &= -A^{-1}B\left(C - B^TA^{-1}B\right)^{-1}, \\
F &= \left[C - B^T A^{-1} B\right]^{-1}.
\end{align}
\end{subequations}

\end{appendix}

\end{document}